\documentclass[aps,prd,10pt,nofootinbib,OliveGeqsecnum,showpacs,showkeys,superscriptaddress,preprintnumbers,altaffilletter,floatfix,twocolumn]{revtex4-2}

\usepackage{amsmath}
\usepackage{multirow}
\usepackage{hyperref}
\usepackage{soul}
\usepackage{graphicx}
\usepackage{dcolumn}
\usepackage{amssymb}
\usepackage{amsfonts}
\usepackage{amsbsy}
\usepackage{color}
\usepackage{rotating}
\usepackage{overpic}
\usepackage{ifxetex}
\usepackage{lipsum}
\usepackage{float}
\usepackage{bbold}
\usepackage{enumerate}
\usepackage{dsfont}
\usepackage{mathrsfs}
\usepackage{mathtools}
\usepackage{tensor}
\usepackage{caption}
\usepackage{subcaption}
\usepackage{ragged2e}
\ifxetex
   \usepackage{polyglossia}
   \setdefaultlanguage{english}
   \usepackage{fontspec}
\else
   \usepackage[english]{babel}
   \usepackage[utf8]{inputenc}
\fi

\definecolor{amaranth}{rgb}{0.90, 0.17, 0.31}
\definecolor{palatinateblue}{rgb}{0.15, 0.23, 0.89}
\definecolor{brightpink}{rgb}{0.86, 0.26, 0.55} 

\hypersetup{ linktoc=all,
    colorlinks, linkcolor={palatinateblue},
    citecolor={brightpink}, urlcolor={amaranth}
}

\graphicspath{ {images/} }

\begin{document}


\title{Dark energy driven by an oscillating generalised axion-like quintessence field}

\author{Mariam Bouhmadi-López}
\email{mariam.bouhmadi[at]ehu.eus}
\affiliation{IKERBASQUE, Basque Foundation for Science, 48011, Bilbao, Spain}
\affiliation{Department of Physics, University of the Basque Country UPV/EHU, P.O. Box 644, 48080 Bilbao, Spain}
\affiliation{EHU Quantum Center, University of the Basque Country UPV/EHU, P.O. Box 644, 48080 Bilbao, Spain}

\author{Carlos G. Boiza}
\email{carlos.garciab[at]ehu.eus}
\affiliation{Department of Physics, University of the Basque Country UPV/EHU, P.O. Box 644, 48080 Bilbao, Spain}
\affiliation{EHU Quantum Center, University of the Basque Country UPV/EHU, P.O. Box 644, 48080 Bilbao, Spain}

\begin{abstract} 

Generalised axion-like scalar fields provide a well-motivated framework for describing the late-time acceleration of the Universe. As the field evolves, it rolls down its potential and, depending on its mass and initial conditions, it may either still be approaching the minimum or already oscillating around it. These two dynamical regimes require distinct treatments of cosmological perturbations. In this work, we perform a detailed analysis of linear cosmological perturbations in the regime where the dark-energy scalar field undergoes coherent oscillations about the minimum of its potential. We show that the standard effective fluid description breaks down in this phase and develop a consistent field-based perturbation framework, which we use to assess the impact of oscillatory dark energy on the growth of cosmic structures.

\end{abstract}

\maketitle

\renewcommand{\tocname}{Index}


\section{Introduction}\label{intro}

Quintessence provides a theoretically well-motivated framework for explaining the
late-time accelerated expansion of the Universe by postulating a canonical scalar
field minimally coupled to gravity \cite{Ratra:1987rm,Wetterich:1987fm}. Unlike a
cosmological constant, whose energy density is strictly time independent,
quintessence is intrinsically dynamical: its evolution is governed by a
self-interaction potential, allowing the equation-of-state parameter to vary with
cosmic time, typically remaining close to, but not exactly equal to, $w=-1$
\cite{Caldwell:1997ii}. As a result, quintessence models offer a
dynamical origin for dark energy and can, at least partially, alleviate conceptual
issues of the $\Lambda$CDM paradigm such as fine-tuning and cosmic coincidence,
while remaining consistent with current observational constraints
\cite{Ratra:1987rm,Wetterich:1987fm,Ferreira:1997hj,Zlatev:1998tr,Liddle:1998xm,Steinhardt:1999nw,Tsujikawa:2013fta}. Beyond their impact on the background expansion
history, such models can also leave observable signatures in the growth of cosmic
structures \cite{PhysRevD.64.103508,Amendola:2015ksp}.

For a canonical, minimally coupled scalar field, several well-defined dynamical
routes exist through which an effective equation of state approaching
$w \to -1$ can be realised. A broad and well-studied class of such models is given
by freezing scenarios, in which the scalar field initially has
$w>-1$ and evolves toward $w \simeq -1$ with $\mathrm{d}w/\mathrm{d}t<0$ as the
kinetic contribution progressively decreases. Within this class, tracking and
scaling solutions allow the scalar field to follow the background evolution during
radiation and matter domination before transitioning to a dark-energy-dominated
phase, thereby helping to alleviate the coincidence problem
\cite{Zlatev:1998tr,Steinhardt:1999nw,Liddle:1998xm}. In most well-studied freezing
models of this type, the de~Sitter--like attractor is realised only in a runaway
regime at $\phi \to \infty$, and the potential does not admit a finite, positive
minimum.

By contrast, in thawing models the scalar field remains frozen by Hubble friction
until relatively late-time, maintaining $w \simeq -1$ over most of the
observationally relevant epoch
\cite{Caldwell:2005tm,Clemson:2008ua,Chiba:2012cb}. Once the field begins to roll,
the equation of state departs from $-1$ with $\mathrm{d}w/\mathrm{d}t>0$, reflecting
the growth of the kinetic energy. In all these cases, the emergence of
$w \simeq -1$ is a natural dynamical outcome of scalar-field evolution rather than
the result of a finely tuned cosmological constant \cite{Tsujikawa:2013fta}.

Observationally, current data tightly constrain departures from $w=-1$ at late-time, but do not uniquely determine the nature of dark energy. 
Taken individually, key probes---Type~Ia supernovae, baryon acoustic oscillations, the cosmic microwave background, and large-scale structure---are broadly consistent with an equation of state close to $w=-1$, and typically allow only limited room for mild evolution within their standalone uncertainties \cite{SupernovaSearchTeam:1998fmf,SupernovaCosmologyProject:1998vns,Brout:2022vxf,Planck:2018vyg,SDSS:2005xqv,2dFGRS:2005yhx,percival2010baryon,SDSS:2003tbn,Reid:2009xm}. 
Joint analyses provide substantially tighter constraints and can, depending on the adopted data combination and systematics, either further restrict late-time evolution or mildly prefer it within time-dependent parameterisations. 
In previous works \cite{Chiang:2025qxg,Bouhmadi-Lopez:2025wxo}, we showed that scalar fields with generalised axion-like potentials achieve statistical parity with $\Lambda$CDM according to Bayesian model comparison and information criteria (see also \cite{Hossain:2025grx}). 
Moreover, recent results from the Dark Energy Spectroscopic Instrument (DESI), including its second data release, have further strengthened interest in dynamical dark energy: combined analyses of DESI baryon acoustic oscillation data with external probes indicate that evolving dark-energy models can fit the data at a level comparable to $\Lambda$CDM, with the strength of any preference depending on the particular combination employed \cite{DESI:2024uvr,DESI:2024lzq,DESI:2024mwx,DESI:2025zgx}. 
These developments motivate a careful and robust treatment of scalar-field dynamics and their imprint on structure formation.

Axion dark energy represents a theoretically appealing realisation of dynamical dark
energy, in which a light pseudoscalar field drives late-time cosmic acceleration
through its potential
\cite{Frieman:1995pm,Kamionkowski:2014zda,Emami:2016mrt,Marsh:2015xka}.
Within this broader class, generalised axion-like models provide a flexible
phenomenological framework capable of producing a rich variety of late-time
behaviours while remaining compatible with current observations
\cite{Hossain:2023lxs,Boiza:2024azh,Boiza:2024fmr,Hossain:2025grx,Chiang:2025qxg,Bouhmadi-Lopez:2025wxo}.
Beyond their role as dark-energy candidates, axion and axion-like scalar fields
have been widely explored in cosmology and particle physics, with applications
ranging from early-dark-energy scenarios connected to the Hubble tension
\cite{Poulin:2018cxd,Kamionkowski:2022pkx} and modifications of the
recombination-era dynamics \cite{Smith:2025uaq}, to inflationary model building
\cite{Freese:1990rb}, primordial black-hole formation mechanisms
\cite{Khlopov:2004sc}, and dark-matter phenomenology in both QCD-inspired and
ultralight regimes \cite{Khlopov:1999tm,Hui:2016ltb}.
Axion-motivated multi-field realisations, such as the string axiverse, further
illustrate how a spectrum of light fields can give rise to rich cosmological
dynamics across cosmic history
\cite{Kamionkowski:2014zda,Emami:2016mrt}, and dedicated numerical simulations
have been developed to characterise the distinctive clustering properties of
ultralight axion dark matter
\cite{Schive:2014dra,Schive:2014hza,Mocz:2019uyd}.

In many well-studied tracking quintessence models, such as inverse power-law
potentials, the scalar field evolves along a runaway direction in field space and
approaches a de~Sitter--like state only in the limit $\phi \to \infty$, without the
presence of a finite positive minimum of the potential \cite{Ratra:1987rm,Zlatev:1998tr,Liddle:1998xm,Steinhardt:1999nw}. By contrast, a distinctive
feature of the class of models considered in this work is that the de~Sitter state is
realised at a finite field value, corresponding to a true positive minimum of the
potential \cite{Boiza:2024azh,Boiza:2024fmr}. As the scalar field relaxes toward this minimum, it may overshoot and
oscillate coherently around it. This behaviour differs qualitatively from the
monotonic approach to de~Sitter characteristic of runaway tracking models and has
important consequences for cosmological perturbations: at the turning points of the
oscillatory motion, the effective fluid description of dark energy becomes ill
defined, even though the underlying scalar-field and metric perturbations remain
perfectly regular.

In this work we focus on the oscillatory regime of generalised axion-like
quintessence and develop a consistent linear perturbation framework that remains
well defined across both oscillatory and non-oscillatory phases. By working
directly with the fundamental scalar-field and metric perturbations, we avoid the
pathologies of the effective fluid approach and obtain a robust description of
the impact of oscillatory dark energy on the growth of cosmic structures.

This paper is organised as follows. In
Section~\ref{sec:gen_axion_summary}, we introduce the generalised axion-like
quintessence model and summarise its background dynamics. In
Section~\ref{perturbations}, we derive the linear cosmological perturbation
equations appropriate for oscillatory scalar-field dynamics. In Section~\ref{numerical results}, we present and analyse numerical solutions of
the perturbation equations and discuss their implications for the growth of
cosmic structures. Finally, in Section~\ref{conclusions}, we summarise
our main results and draw our conclusions.

\section{Generalised axion-like quintessence with matter and radiation} \label{sec:gen_axion_summary}

In this section we summarise the quintessence model introduced in \cite{Hossain:2023lxs,Boiza:2024azh}, which generalises axion-like scalar-field potentials to
describe the late-time accelerated expansion of the Universe. The model is formulated
within a spatially flat Friedmann--Lemaître--Robertson--Walker (FLRW) cosmology and
includes standard radiation and pressureless matter components, which dominate the
cosmic expansion at early-time and provide the background against which the
scalar-field dynamics unfold.

\subsection{Background dynamics}
\label{subsec:background_dynamics}

We consider a canonical scalar field $\phi$ minimally coupled to gravity, evolving in a
spatially flat FLRW spacetime with line element
\begin{equation}
\mathrm{d}s^2 = -\mathrm{d}t^2 + a^2(t)\,\delta_{ij}\,\mathrm{d}x^i\mathrm{d}x^j ,
\end{equation}
where $a(t)$ is the cosmological scale factor, and coexisting with standard radiation
and pressureless matter.

The action of the system is
\begin{equation}
S = \int \mathrm{d}^4x \sqrt{-g}
\left[
\frac{1}{2\kappa^2} R
- \frac{1}{2} g^{\mu\nu}\partial_\mu\phi\,\partial_\nu\phi
- V(\phi)
\right]
+ S_{\rm r} + S_{\rm m},
\end{equation}
where $\kappa^2 \equiv 8\pi G$, and $S_{\rm r}$ and $S_{\rm m}$ denote the actions for the
radiation and matter components, respectively.

Variation of the action with respect to the scalar field yields the homogeneous
Klein--Gordon equation,
\begin{equation}
\label{eq:KG}
\ddot{\phi} + 3H\dot{\phi} + V_{,\phi} = 0 ,
\end{equation}
where $V_{,\phi}\equiv \mathrm{d}V/\mathrm{d}\phi$, and the Hubble parameter
$H\equiv \dot{a}/a$ is defined in terms of the scale factor.

The expansion of the Universe is governed by the Friedmann equations,
\begin{eqnarray}
&&H^2 = \frac{\kappa^2}{3}
\left(\rho_{\rm r} + \rho_{\rm m} + \rho_\phi\right), \label{eq:Friedmann1}\\
&&\dot{H} = -\frac{\kappa^2}{2}
\left(\frac{4}{3}\rho_{\rm r} + \rho_{\rm m} + \rho_\phi + p_\phi\right), \label{eq:Friedmann2}
\end{eqnarray}
where the scalar field behaves as an effective perfect fluid with energy density and
pressure
\begin{equation}
\rho_\phi = \frac{1}{2}\dot{\phi}^2 + V(\phi),
\qquad
p_\phi = \frac{1}{2}\dot{\phi}^2 - V(\phi).
\end{equation}

Radiation and pressureless matter satisfy the standard continuity equations,
\begin{equation}
\dot{\rho}_{\rm r} + 4H\rho_{\rm r} = 0,
\qquad
\dot{\rho}_{\rm m} + 3H\rho_{\rm m} = 0,
\end{equation}
so that radiation dominates the expansion at early-time, followed by a matter-dominated
era, while the scalar field becomes dynamically relevant only at late-time.

This framework provides the minimal and well-established setting for quintessence models
of dark energy, within which the specific properties of the scalar-field potential
determine the detailed late-time cosmological dynamics.

\subsection{Generalised axion-like potential and effective mass}
\label{subsec:axion_potential}

The dynamics of the scalar field are governed by a generalised axion-like potential of
the form
\begin{equation}
\label{eq:gen_axion_potential}
V(\phi) = V_0
\left[
1 - \cos\!\left(\frac{\phi}{\eta}\right)
\right]^{-n},
\quad V_0>0,\quad n>0,\quad \eta>0 .
\end{equation}
Here $V_0$ sets the overall energy scale of the potential and $\eta$ is a constant with
dimensions of mass. The conditions $V_0>0$, $n>0$, and $\eta>0$ ensure that the potential
is positive definite and admits a well-defined minimum. This potential generalises the
standard axion-like form commonly considered in quintessence and dark-energy models
\cite{Frieman:1995pm,Kamionkowski:2014zda,Emami:2016mrt,Marsh:2015xka}, and is treated here as an effective phenomenological description of the
scalar-field self-interactions.

\begin{figure}[t]
  \centering
  \includegraphics[width=0.48\textwidth]{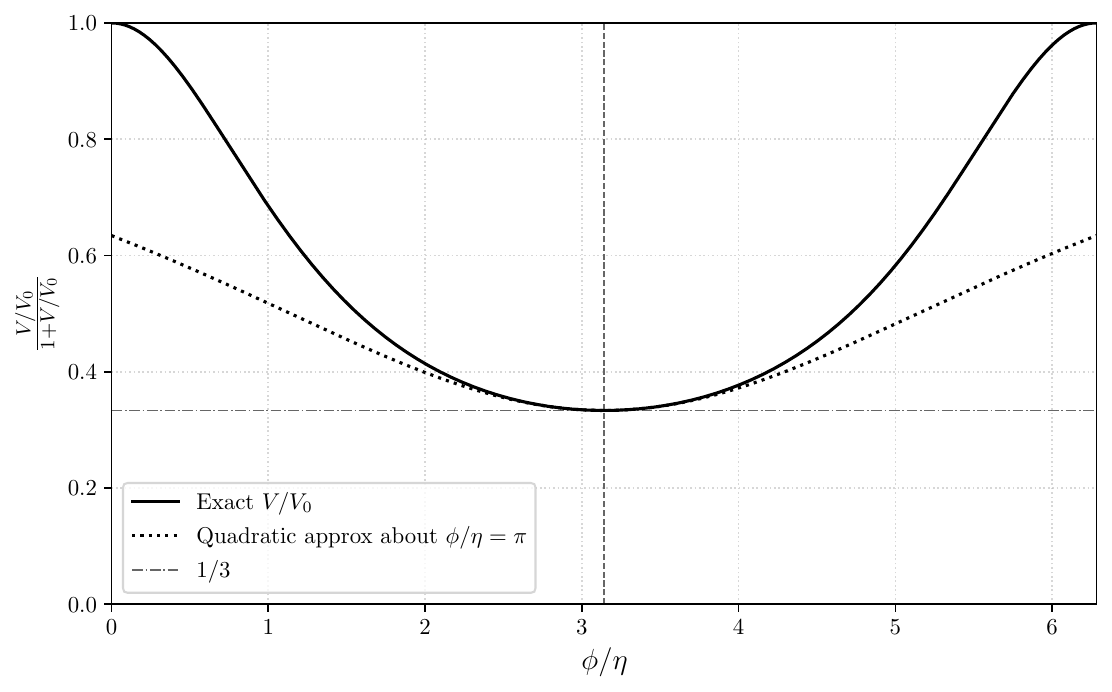}
  \caption{\justifying{Exact generalised axion-like potential
  $V(\phi)/V_0=\left[1-\cos(\phi/\eta)\right]^{-n}$ (solid line) and its quadratic
  approximation around the minimum at $\phi/\eta=\pi$ (dotted line), shown as functions
  of the dimensionless field variable $\phi/\eta$ for the representative case $n=1$.
  The potential is displayed in the compactified form $(V/V_0)/(1+V/V_0)$ in order to
  avoid divergences at the extrema. In this representation, the minimum of the
  potential corresponds to $(V/V_0)/(1+V/V_0)=1/3$.}}
  \label{fig:potential_exact_vs_quadratic}
\end{figure}

For suitable parameter choices, the potential admits a prolonged dynamical regime in
which the scalar field tracks the dominant background component during radiation- and
matter-dominated eras, before eventually driving late-time cosmic acceleration, as
discussed in \cite{Boiza:2024azh}. The overall shape of the potential and its
behaviour near the minimum are illustrated in
Fig.~\ref{fig:potential_exact_vs_quadratic} for the representative case $n=1$.

The potential~\eqref{eq:gen_axion_potential} possesses a minimum at
\begin{equation}
\phi_{\rm min} = \pi \eta ,
\end{equation}
where the first derivative of the potential vanishes. The value of the potential at the
minimum is given by
\begin{equation}
\label{eq:Vmin_def}
V_{\rm min} \equiv V(\phi_{\rm min}) = 2^{-n}\,V_0.
\end{equation}
For the illustrative case $n=1$ shown in Fig.~\ref{fig:potential_exact_vs_quadratic},
this corresponds to $V_{\rm min}/V_0 = 1/2$.

Expanding the potential around the minimum yields
\begin{equation}
V(\phi) \simeq V_{\rm min}
+ \frac{1}{2} m_{\rm eff}^2 \, (\phi-\phi_{\rm min})^2
+ \mathcal{O}\!\left[(\phi-\phi_{\rm min})^3\right],
\end{equation}
which defines the effective mass of the scalar field,
\begin{equation}
\label{eq:meff_def}
m_{\rm eff}^2
\equiv V_{,\phi\phi}\big|_{\phi=\phi_{\rm min}}
= \frac{n}{2\eta^2}\,V_{\rm min}.
\end{equation}

Such a quadratic expansion around the minimum and the associated effective mass are
standard in the analysis of coherently oscillating scalar fields in cosmology
\cite{Marsh:2015xka}. Equation~\eqref{eq:meff_def} shows explicitly that the
effective mass is determined by the curvature of the potential at its minimum and by
the vacuum energy scale $V_{\rm min}$. In particular, increasing $n$ or decreasing
$\eta$ leads to a larger $m_{\rm eff}$, thereby favouring an earlier onset of oscillatory
behaviour once the field reaches the vicinity of the minimum. Near $\phi_{\rm min}$, the quadratic approximation provides an accurate description of
the local dynamics of the scalar field and captures the essential features of the
oscillatory regime.

\subsection{Tracking behaviour, late-time acceleration, and oscillatory criterion}
\label{subsec:tracking_late_time}

The cosmological evolution of the generalised axion-like scalar field typically
proceeds through an early-time tracking regime followed by a late-time phase in which
the field becomes dominant and drives cosmic acceleration. Importantly, the early-time
tracking behaviour of this model is \emph{not} a scaling solution: although the
scalar-field energy density remains subdominant and evolves in a manner largely
insensitive to initial conditions, the scalar-field equation of state does not, in
general, coincide with that of the dominant background component.

During matter domination, the tracker solution is well approximated by
\cite{Boiza:2024azh,Boiza:2024fmr}
\begin{equation}
\label{eq:wphi_tracker_matter}
w_\phi \simeq -\frac{1}{1+n},
\qquad (\text{matter domination}),
\end{equation}
so that the scalar-field equation of state depends explicitly on the potential parameter
$n$. In particular, for $n=1$ one obtains $w_\phi\simeq -1/2$, while smaller values of $n$
yield $w_\phi$ closer to $-1$, making the background evolution increasingly similar to
that of a cosmological constant already during the tracking phase. In this regime, the
scalar-field energy density decreases more slowly than that of matter, allowing
$\Omega_\phi$ to grow with time and naturally triggering a transition to dark-energy
domination at late-time.

As the field approaches the minimum of the potential, the dynamics become controlled by
the local curvature scale, characterised by the effective mass $m_{\rm eff}$ defined in
Eq.~\eqref{eq:meff_def}. Once the field is sufficiently close to the minimum
$\phi_{\rm min}=\pi\eta$, the field displacement
$\Delta_\phi\equiv\phi-\phi_{\rm min}$ satisfies, to leading order,
\begin{equation}
\ddot{\Delta}_\phi + 3H\dot{\Delta}_\phi + m_{\rm eff}^2\,\Delta_\phi \simeq 0,
\end{equation}
which corresponds to a damped harmonic oscillator with Hubble friction. If $H$ is
approximately constant over an oscillation period,
the ansatz $\Delta_\phi\propto e^{\lambda t}$ gives
\begin{equation}
\lambda_\pm=-\frac{3H}{2}\pm\sqrt{\left(\frac{3H}{2}\right)^2-m_{\rm eff}^2}\, .
\end{equation}
The motion is underdamped (and therefore oscillatory) only when the discriminant is
negative, i.e.
\begin{equation}
m_{\rm eff}>\frac{3}{2}H .
\end{equation}
Since $H(t)$ varies slowly at late-time and the transition from overdamped to
underdamped behaviour is gradual, it is useful to express this as an order-of-magnitude
diagnostic in terms of the present-day expansion rate $H_0$. This motivates
\begin{equation}
\label{eq:oscillation_criterion}
\begin{aligned}
m_{\rm eff} \gtrsim H_0
&\Rightarrow \text{oscillations are possible by today},\\
m_{\rm eff} \lesssim H_0
&\Rightarrow \text{the field remains overdamped},
\end{aligned}
\end{equation}
where the second line follows because in the overdamped regime the solution is a sum of
two decaying exponentials rather than a sinusoid.

A useful corollary of Eq.~\eqref{eq:oscillation_criterion} is that it implies an
order-of-magnitude critical scale $\eta_{\rm crit}$ separating models that have
already entered coherent oscillations by the time the field reaches the vicinity
of the minimum from those that remain overdamped. Using
Eq.~\eqref{eq:meff_def}, the oscillation condition
$m_{\rm eff}\sim \mathcal{O}(H)$ can be written as
$\eta\sim \sqrt{nV_{\rm min}}/H$.
If, in addition, the field reaches the minimum during the onset of vacuum-energy
domination so that $3H^2\simeq \kappa^2 V_{\rm min}$, the dependence on $V_{\rm min}$
cancels and one finds $\eta_{\rm crit}\sim \sqrt{n}\,\kappa^{-1}$.
In this approximation, $\eta_{\rm crit}$ is therefore set primarily by the
gravitational scale $\kappa^{-1}$ (up to an $\mathcal{O}(1)$ factor depending on $n$)
and is only weakly sensitive to late-time cosmological parameters such as $H_0$ and
$\Omega_{m0}$.
More generally, if the minimum is reached when matter, radiation, or the scalar
kinetic term still contribute non-negligibly to the expansion rate,
$3H^2\simeq \kappa^2\left(V_{\rm min}+K+\rho_m+\rho_r\right)$, the cancellation is only
approximate and $\eta_{\rm crit}$ acquires mild corrections of relative size
$\sim[1+K/V_{\rm min}+(\rho_m+\rho_r)/V_{\rm min}]^{-1/2}$.
In the late-time accelerating regime relevant for the oscillatory solutions studied
here, one typically has $w_\phi\simeq -1$, so that $K\ll V_{\rm min}$ and matter is
already subdominant, keeping the dependence of $\eta_{\rm crit}$ on these details
modest in practice.

\begin{figure*}[t]
  \centering
  \begin{subfigure}[t]{0.48\textwidth}
    \centering
    \includegraphics[width=\textwidth]{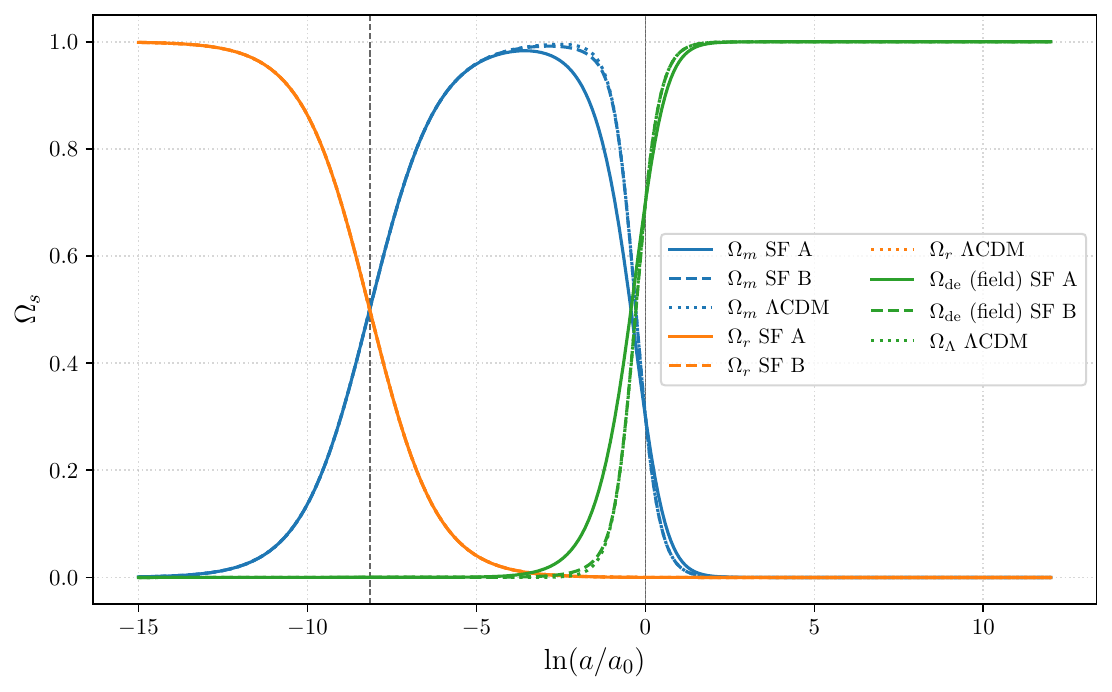}
    \caption{\justifying{Evolution of the density parameters $\Omega_m$, $\Omega_r$,
    and $\Omega_\phi$ for two representative generalised axion-like quintessence models,
    corresponding to $\eta=1$ (non-oscillatory) and $\eta=0.1$ (oscillatory).
    During the tracking phase the scalar-field density fraction grows relative to
    matter, eventually dominating the cosmic expansion and driving late-time
    acceleration.}}
    \label{fig:background_omegas}
  \end{subfigure}
  \hfill
  \begin{subfigure}[t]{0.48\textwidth}
    \centering
    \includegraphics[width=\textwidth]{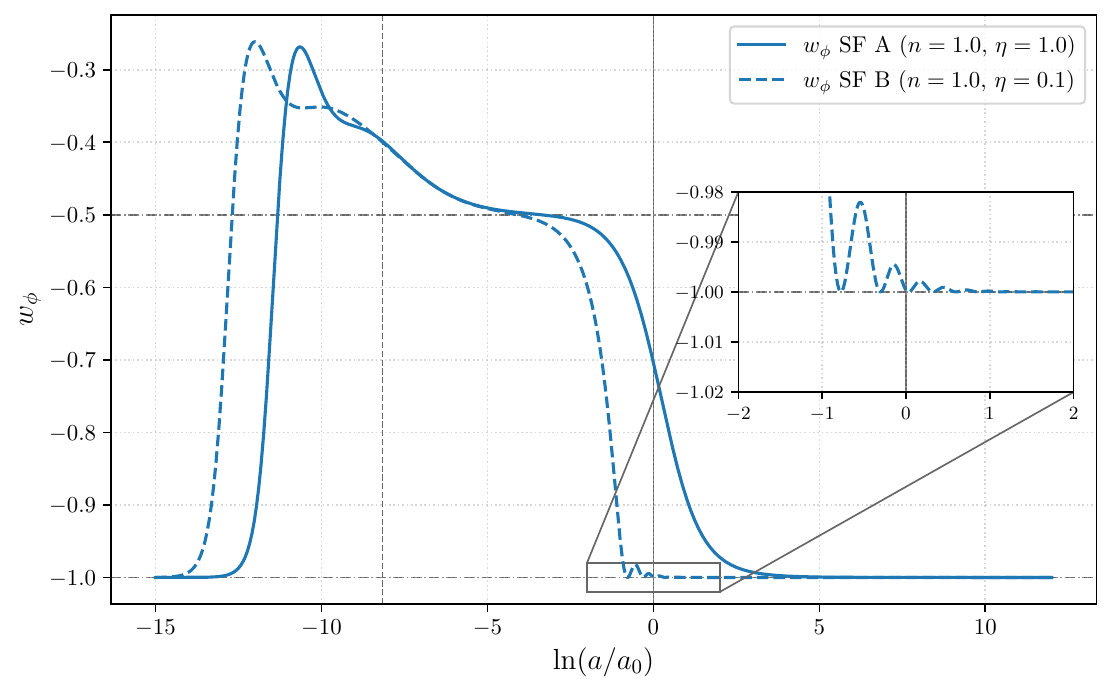}
    \caption{\justifying{Evolution of the density parameters $\Omega_m$, $\Omega_r$, and $\Omega_\phi$ for the two benchmark models shown throughout this work: SF A ($\eta=1$, non-oscillatory) and SF B ($\eta=0.1$, oscillatory). During the tracking phase the scalar-field density fraction grows relative to matter, eventually dominating the cosmic expansion and driving late-time acceleration.}}
    \label{fig:background_wphi}
  \end{subfigure}
  \caption{\justifying{Background evolution of generalised axion-like quintessence. The left panel shows the transition from radiation and matter domination to a scalar-field-dominated accelerated phase, while the right panel shows the associated evolution of the scalar-field equation of state. The comparison highlights the qualitative difference between the non-oscillatory benchmark SF A ($\eta=1$) and the oscillatory benchmark SF B ($\eta=0.1$) at late-time.}}
  \label{fig:background_evolution}
\end{figure*}

The background evolution for representative models is illustrated in
Fig.~\ref{fig:background_evolution}. The left panel shows the evolution of the density
parameters, including the growth of $\Omega_\phi$ during the tracking phase and the
transition to scalar-field domination at late-time. The right panel displays the
corresponding evolution of the scalar-field equation-of-state parameter, highlighting
both the non-scaling nature of the tracking regime and the late-time approach towards
$w_\phi\simeq -1$ as the field settles near the potential minimum. \footnote{All numerical results and figures in this section are computed for $n=1$.
The background evolution is obtained by integrating the homogeneous equations
with initial conditions $\phi_i/\eta = 10^{-4}$ and $\dot{\phi}_i=0$.
Owing to the presence of an early-time tracking regime, the late-time evolution
of the scalar field is largely insensitive to the precise choice of these initial
conditions; the values adopted here are therefore chosen for numerical stability
and convenience.
We fix the present-day matter density parameter to $\Omega_{m0}=0.3$.
The present-day Hubble parameter is fixed to
$H_0=69.5705~\mathrm{km\,s^{-1}\,Mpc^{-1}}$, corresponding to
$h\equiv H_0/(100~\mathrm{km\,s^{-1}\,Mpc^{-1}})$, a value that lies between
the cosmic-microwave-background–inferred determination from \cite{Planck:2018vyg}
and recent Type~Ia supernova measurements \cite{Brout:2022vxf}.
The radiation density is set using
$\Omega_{r0}=\Omega_{m0}/(1+z_{\rm eq})$, with
$z_{\rm eq}=2.5\times10^4\,(\Omega_{m0}h^2)\,(T_{\rm CMB}/2.7)^{-4}$ and
$T_{\rm CMB}=2.7255~\mathrm{K}$ \cite{Eisenstein:1997ik,Fixsen:2009ug}.
The parameter $V_0$ is chosen such that the closure relation
$\Omega_{m0}+\Omega_{r0}+\Omega_{\phi0}=1$ is satisfied.
}

It is worth emphasising that the presence of a late-time oscillatory regime constitutes
a distinctive feature of the generalised axion-like potential considered here. In many
well-studied tracking quintessence models, such as inverse power-law potentials, late-time
cosmic acceleration arises from an asymptotic evolution in which the scalar field rolls
monotonically towards $\phi\to\infty$, with no minimum and therefore no oscillatory
behaviour \cite{Zlatev:1998tr,Steinhardt:1999nw,Liddle:1998xm}. In those cases, the tracking
solution smoothly connects to a slow-roll–like dark-energy phase, and the fluid
description of perturbations remains well defined at all times.

By contrast, the generalised axion-like potential admits a finite minimum that can be
reached at late-time, opening the possibility of coherent oscillations precisely during
the dark-energy-dominated epoch. This feature places the model outside the standard
quintessence paradigm and necessitates a careful reassessment of both background and
perturbative dynamics in the late Universe.

The distinction between oscillatory and non-oscillatory regimes has important
implications for the treatment of cosmological perturbations. In the absence of
oscillations ($m_{\rm eff}\lesssim H_0$), the scalar field evolves slowly and can be
consistently described as an effective dark-energy fluid, allowing the use of a
multi-fluid perturbation framework in which both the equation-of-state parameter
$w_\phi$ and the adiabatic sound speed remain finite and slowly varying. In this case,
the modified clustering properties of dark energy generically lead to a suppression
of matter perturbations, with potentially observable consequences for the growth of
structure and related cosmological probes.

Once coherent oscillations are present ($m_{\rm eff}\gtrsim H_0$), this description
breaks down. Although the instantaneous equation of state
$w_\phi=p_\phi/\rho_\phi$ remains well defined throughout the evolution, the adiabatic
sound speed $c_{a\phi}^2 \equiv \dot p_\phi/{\dot\rho_\phi}$ diverges whenever
$w_\phi=-1$, which necessarily occurs at each turning point of the oscillatory motion
around the potential minimum. As a result, the adiabatic sound speed entering the fluid
perturbation equations becomes ill defined, rendering the standard multi-fluid
description inconsistent in the oscillatory regime \cite{Boiza:2024fmr}. 
At the background level, however, the oscillatory solutions are characterised by an
equation of state that remains very close to $w_\phi\simeq -1$ once the scalar field
becomes dynamically relevant. Consequently, the scalar field behaves effectively as a
cosmological constant from the perspective of structure formation, strongly suppressing
the impact of dark-energy perturbations on the growth of matter fluctuations.

These two behaviours are explicitly illustrated by the benchmark models shown in
Figs.~\ref{fig:background_omegas} and~\ref{fig:background_wphi}. For SF A ($\eta=1$, measured in units of the gravitational scale $\kappa^{-1}$),
the effective mass satisfies $m_{\rm eff}\lesssim H_0$, and the field has not completed a
full oscillation by the present epoch. In this case the scalar field remains in a
non-oscillatory regime with $w_\phi$ appreciably different from $-1$ for an extended
period, during which it contributes non-negligibly to the total energy density and
modifies the expansion history, leading to a potentially observable suppression of
matter clustering. By contrast, for SF B ($\eta=0.1$) one finds $m_{\rm eff}\gtrsim H_0$,
so that coherent oscillations are already present today. The scalar field then settles
into an evolution with $w_\phi\simeq -1$ on average, and both the background dynamics and
the associated matter perturbations closely mimic those of $\Lambda$CDM, making
deviations in late-time clustering significantly less apparent.

Nevertheless, oscillatory solutions cannot be disregarded on phenomenological grounds.
Indeed, our analysis of observational data in \cite{Chiang:2025qxg} revealed a
bimodal structure in the allowed parameter space, encompassing both non-oscillatory
models—such as the $\eta=1$ case—and oscillatory models, exemplified by $\eta=0.1$.

These considerations motivate a unified perturbative treatment that remains valid
across both regimes and does not rely on a fluid approximation. In the following
section we therefore develop the linear perturbation equations directly at the level
of the scalar field and metric fluctuations, enabling a consistent and robust
assessment of generalised axion-like quintessence models across the full parameter
space.

\section{Cosmological perturbations with oscillating scalar fields}\label{perturbations}

In this section we develop the linear cosmological perturbation theory for
generalised axion-like quintessence models in the presence of coherent
scalar-field oscillations.
As discussed in Sec.~\ref{subsec:tracking_late_time}, the late-time dynamics of
these models can enter an oscillatory regime in which the usual effective
fluid description of dark energy breaks down.
In particular, although the background evolution remains well behaved, the
adiabatic sound speed associated with a fluid approximation becomes ill
defined whenever the scalar-field equation of state approaches
$w_\phi=-1$, which inevitably occurs during each oscillation around the
minimum of the potential.

To overcome this limitation, we formulate the perturbation equations directly
in terms of the fundamental scalar-field and metric fluctuations, without
relying on a multi-fluid description of dark energy.
This approach remains well defined in both oscillatory and non-oscillatory
regimes and allows for a consistent treatment of perturbations across the
entire parameter space of the model.

In this section we focus on the theoretical formulation of the perturbation
equations and on their general properties.
A detailed numerical analysis of scalar-field and matter perturbations, as
well as their impact on observable quantities such as the matter power
spectrum and the growth-rate parameter $f\sigma_8$, is presented in the
following section.
Further technical aspects of the breakdown of the fluid description in the
oscillatory regime, together with additional numerical diagnostics, are
discussed in Appendix~\ref{app:fluid_pathology_metric_regular}.

Cosmological perturbations have been extensively studied in the literature in a
variety of contexts; see, for instance, \cite{Mukhanov:1990me,Ma:1995ey,Brandenberger:2003vk,Malik:2008im,Malquarti:2002iu,Bassett:2005xm,Wang:2009azb,baumann}
for general reviews and formulations.
In particular, \cite{Albarran:2016mdu} presents a closely related treatment of perturbations, employing a notation and gauge choice similar to the one
adopted in this work.

\subsection{Linear perturbations and gauge choice}
\label{subsec:pert_general}

We study scalar perturbations around a spatially flat FLRW background in the
Newtonian gauge. The perturbed line element is
\begin{equation}
\label{eq:newtonian_metric}
\mathrm{d}s^2
=
a^2(\tau)\Big[
-(1+2\Phi)\,\mathrm{d}\tau^2
+
(1-2\Psi)\,\delta_{ij}\,\mathrm{d}x^i\mathrm{d}x^j
\Big],
\end{equation}
where $\tau$ denotes conformal time, related to cosmic time by
$\mathrm{d}\tau=\mathrm{d}t/a$, and $\Phi(\tau,\mathbf{x})$ and
$\Psi(\tau,\mathbf{x})$ are the Bardeen potentials \cite{Bardeen:1980kt}. Throughout this work we assume
negligible anisotropic stress for the matter and radiation components, so that
\begin{equation}
\label{eq:phi_equals_psi}
\Phi=\Psi.
\end{equation}
We work in Fourier space, where each perturbation variable $X(\tau,\mathbf{x})$
is decomposed as $X(\tau,\mathbf{x})=\int \mathrm{d}^3k\,X(\tau,\mathbf{k})
\,e^{i\mathbf{k}\cdot\mathbf{x}}$ and the Laplacian acts as
$\nabla^2\rightarrow -k^2$.

The scalar field is perturbed as
\begin{equation}
\phi(\tau,\mathbf{x})=\bar{\phi}(\tau)+\delta\phi(\tau,\mathbf{x}),
\end{equation}
where an overbar denotes the homogeneous background. The background field obeys
the Klein--Gordon equation
\begin{equation}
\label{eq:bg_KG_conformal}
\bar{\phi}''+2\mathcal{H}\bar{\phi}'+a^2 V_{,\phi}(\bar{\phi})=0,
\end{equation}
with $\mathcal{H}\equiv a'/a$ the conformal Hubble rate and a prime denoting
$\mathrm{d}/\mathrm{d}\tau$.

\subsubsection*{Einstein equations at linear order}

The linearised Einstein equations in the Newtonian gauge yield the usual
constraint and evolution equations for $\Psi$ \cite{baumann}. The $(0,0)$ component gives
\begin{equation}
\label{eq:poisson_like}
k^2\Psi+3\mathcal{H}\left(\Psi'+\mathcal{H}\Psi\right)
=
-4\pi G a^2\,\delta\rho,
\end{equation}
while the $(0,i)$ component implies
\begin{equation}
\label{eq:momentum_constraint}
\Psi'+\mathcal{H}\Psi
=
-4\pi G a^2\,(\bar{\rho}+\bar{p})\,v,
\end{equation}
where $\delta\rho$ is the total energy-density perturbation and $v$ is the total
velocity potential defined through $\delta T^{0}{}_{i}=(\bar{\rho}+\bar{p})
\,\partial_i v$. Finally, the traceless $(i,j)$ equation enforces
Eq.~\eqref{eq:phi_equals_psi} in the absence of anisotropic stress, and the trace
part provides the dynamical equation
\begin{equation}
\label{eq:psi_evolution}
\Psi''+3\mathcal{H}\Psi'+\left(2\mathcal{H}'+\mathcal{H}^2\right)\Psi
=
4\pi G a^2\,\delta p,
\end{equation}
with $\delta p$ the total pressure perturbation.

The total perturbed sources are the sums over matter, radiation, and the scalar
field:
\begin{eqnarray}
&&\delta\rho = \delta\rho_{\rm m}+\delta\rho_{\rm r}+\delta\rho_\phi,\\
&&\delta p = \delta p_{\rm r}+\delta p_\phi,\\
&&(\bar{\rho}+\bar{p})\,v = \sum_A(\bar{\rho}_A+\bar{p}_A)\,v_A.
\end{eqnarray}
where $A\in\{{\rm m},{\rm r},\phi\}$ and $\bar{p}_{\rm m}=0$, $\bar{p}_{\rm r}=\bar{\rho}_{\rm r}/3$.

\subsubsection*{Scalar-field perturbation equation}

The perturbation $\delta\phi$ is governed by the linearised Klein--Gordon equation
in the Newtonian gauge \cite{Malquarti:2002iu,Bassett:2005xm,Wang:2009azb},
\begin{equation}
\label{eq:delta_phi_KG}
\delta\phi''+2\mathcal{H}\delta\phi'
+\left(k^2+a^2 V_{,\phi\phi}(\bar{\phi})\right)\delta\phi
=
4\bar{\phi}'\Psi'-2a^2 V_{,\phi}(\bar{\phi})\,\Psi,
\end{equation}
which remains well defined even when $\bar{\phi}'$ crosses zero during coherent
oscillations.

For completeness, it is useful to recall the scalar-field contributions to the
perturbed energy-momentum tensor in this gauge:
\begin{eqnarray}
\label{eq:scalar_sources}
&&\delta\rho_\phi
=
\frac{1}{a^2}\left(\bar{\phi}'\,\delta\phi'-\bar{\phi}'^{\,2}\Psi\right)
+V_{,\phi}(\bar{\phi})\,\delta\phi,\\
&&\delta p_\phi
=
\frac{1}{a^2}\left(\bar{\phi}'\,\delta\phi'-\bar{\phi}'^{\,2}\Psi\right)
-V_{,\phi}(\bar{\phi})\,\delta\phi,\\
&&(\bar{\rho}_\phi+\bar{p}_\phi)\,v_\phi
=
-\frac{\bar{\phi}'}{a^2}\,\delta\phi,
\end{eqnarray}
where the last relation follows from $\delta T^{0}{}_{i\,(\phi)}
=-\bar{\phi}'\,\partial_i\delta\phi/a^2$.
These expressions allow us to close the Einstein system
Eqs.~\eqref{eq:poisson_like}--\eqref{eq:psi_evolution} in terms of the dynamical
variables $(\Psi,\delta\phi)$ and the matter/radiation perturbations.

\subsubsection*{Matter and radiation perturbations}

We model radiation and pressureless matter as perfect fluids obeying
$\nabla_\mu T^{\mu}{}_{\nu\,(A)}=0$ at linear order. In the Newtonian gauge, their
continuity and Euler equations in Fourier space can be written as \cite{Albarran:2016mdu}
\begin{align}
\label{eq:rad_pert}
\delta_{\rm r}'-\frac{4}{3}\left(k^2 v_{\rm r}+3\Psi'\right)&=0,
&
v_{\rm r}'+\frac{1}{4}\delta_{\rm r}+\Psi&=0,\\
\label{eq:mat_pert}
\delta_{\rm m}'-\left(k^2 v_{\rm m}+3\Psi'\right)&=0,
&
v_{\rm m}'+\mathcal{H}v_{\rm m}+\Psi&=0,
\end{align}
where $\delta_A\equiv \delta\rho_A/\bar{\rho}_A$ and the velocity potentials are
defined via $\delta T^{0}{}_{i\,(A)}=(\bar{\rho}_A+\bar{p}_A)\partial_i v_A$.

Equations~\eqref{eq:poisson_like}--\eqref{eq:mat_pert}, together with the background
system, provide a closed set of linear perturbation equations once a choice of
initial conditions is specified. In the next subsection we explain why, in the
oscillatory regime, treating the scalar field as an effective fluid leads to a
breakdown of the standard multi-fluid formulation, and we motivate the field-based
approach adopted here.

\subsection{Breakdown of the multi-fluid description in the oscillatory regime}
\label{subsec:fluid_breakdown}

In the non-oscillatory regime of quintessence, where the scalar field evolves
monotonically and remains slow rolling at late-time, it is often convenient to
describe the scalar field as an effective dark-energy fluid. In this picture, the
field is characterised by an equation-of-state parameter
$w_\phi \equiv p_\phi/\rho_\phi$ together with an adiabatic sound speed
$c_{a\phi}^2 \equiv \dot p_\phi/\dot\rho_\phi$ and an effective (rest-frame) sound
speed $c_{s\phi}^2=1$. This framework was successfully employed in our previous
work \cite{Boiza:2024fmr} to study linear perturbations in the tracking and
non-oscillatory regimes.

However, once the scalar field enters a regime of coherent oscillations around the
minimum of its potential, the effective fluid description ceases to be valid.
This breakdown is not associated with any pathology of the underlying field
theory, but rather with the failure of the fluid variables used to describe it.

\subsubsection*{Divergence of the adiabatic sound speed}

The key quantity responsible for the breakdown of the fluid picture is the
adiabatic sound speed,
\begin{equation}
\label{eq:ca_def}
c_{a\phi}^2 \equiv \frac{\dot p_\phi}{\dot\rho_\phi}.
\end{equation}
For a canonical scalar field,
\begin{equation}
\rho_\phi = \frac{1}{2}\dot{\phi}^2 + V(\phi),
\qquad
p_\phi = \frac{1}{2}\dot{\phi}^2 - V(\phi),
\end{equation}
so that
\begin{equation}
\dot\rho_\phi = \dot{\phi}\left(\ddot{\phi}+V_{,\phi}\right),
\qquad
\dot p_\phi = \dot{\phi}\left(\ddot{\phi}-V_{,\phi}\right).
\end{equation}
Using the background Klein--Gordon equation
$\ddot{\phi}+3H\dot{\phi}+V_{,\phi}=0$, one finds
\begin{equation}
\label{eq:ca_explicit}
c_{a\phi}^2
=
1+\frac{2V_{,\phi}}{3H\dot{\phi}}.
\end{equation}

In the oscillatory regime, the scalar field crosses the minimum of the potential
repeatedly. At each turning point of the oscillation,
\begin{equation}
\dot{\phi}=0,
\qquad
w_\phi = -1,
\end{equation}
while $V_{,\phi}$ remains finite. As a result, the adiabatic sound speed
$c_{a\phi}^2$ diverges at each oscillation. Since $c_{a\phi}^2$ enters explicitly
into the fluid perturbation equations, this divergence renders the multi-fluid
system ill defined.

Importantly, this divergence does \emph{not} signal a physical instability of the
underlying theory. Rather, it reflects a breakdown of the effective fluid
variables used to describe the scalar field. In particular, while the density
perturbation $\delta_\phi$ remains finite throughout the evolution, the velocity
potential $v_\phi$ becomes ill defined whenever $\dot{\phi}=0$, as it
is proportional to $\delta\phi/\dot{\phi}$. These points correspond to the
turning points of the oscillatory motion, where the scalar field momentarily
behaves as a pure cosmological constant with $w_\phi=-1$. The
fundamental field perturbation $\delta\phi$, as well as the metric perturbations,
remain perfectly regular across these crossings (see Appendix~\ref{app:fluid_pathology_metric_regular}); only the mapping from field
variables to effective fluid variables fails.

\subsubsection*{Failure of the effective fluid perturbation equations}

In the fluid approach, scalar-field perturbations are usually evolved through
equations of the form \cite{Boiza:2024fmr}
\begin{align}
\delta_\phi' &=
-3\mathcal{H}\left(c_{s\phi}^2-w_\phi\right)\delta_\phi
+(1+w_\phi)\left(k^2 v_\phi+3\Psi'\right)
\nonumber\\
&\quad
+9\mathcal{H}^2\left(c_{s\phi}^2-c_{a\phi}^2\right)(1+w_\phi)v_\phi,
\\
v_\phi' &=
-\mathcal{H}(1-3c_{s\phi}^2)v_\phi
-\frac{c_{s\phi}^2}{1+w_\phi}\delta_\phi
-\Psi,
\end{align}
where primes denote derivatives with respect to conformal time.
In the oscillatory regime, these equations become ill defined due to:
\begin{itemize}
\item divergences in $c_{a\phi}^2$,
\item the vanishing of $1+w_\phi$ at turning points.
\end{itemize}

As a consequence, numerical integration of the fluid equations becomes unstable
and physically unreliable, even though the true perturbations of the scalar
field remain finite and well behaved.

\subsubsection*{Field-based perturbation approach}

To consistently describe perturbations in the oscillatory regime, one must abandon
the effective fluid description of dark energy and work directly with the
fundamental perturbation variables of the theory. In practice, this means evolving:
\begin{itemize}
\item the metric perturbation $\Psi$, and
\item the scalar-field perturbation $\delta\phi$,
\end{itemize}
using the linearised Einstein equations
(Eqs.~\eqref{eq:poisson_like}--\eqref{eq:psi_evolution})
together with the perturbed Klein--Gordon equation
\eqref{eq:delta_phi_KG}.

This formulation remains fully regular even when $\dot{\phi}=0$ and is therefore
valid across both oscillatory and non-oscillatory regimes. Matter and radiation
continue to be treated as standard perfect fluids, while the scalar field is
handled as a fundamental degree of freedom rather than an effective fluid.

In the next subsection, we specify the initial conditions used to solve the
perturbation equations numerically, paying particular attention to the treatment
of super-Hubble modes and the matching to adiabatic initial conditions at early-time.

\subsection{Initial conditions}
\label{inicond}

The numerical integration of the linear perturbation equations requires the
specification of initial conditions at an early epoch. We choose the initial
time such that
\begin{equation}
\ln(a_i/a_0) = -15 ,
\end{equation}
which lies deep in the radiation-dominated era. At this time the radiation
component dominates the total energy density, so that
\begin{equation}
\rho_i \simeq \rho_{{\rm r}i},
\qquad
w_{{\rm eff},i} \simeq \frac{1}{3},
\end{equation}
and the contribution of matter and of the scalar field to the background
evolution is negligible. All Fourier modes considered in this work satisfy
\begin{equation}
k \ll \mathcal{H}_i ,
\end{equation}
and therefore lie well outside the Hubble radius.

In this super-Hubble regime, the growing solution of the perturbed Einstein
equations corresponds to a constant Newtonian potential \cite{baumann}. We select this mode by
imposing
\begin{equation}
\Psi'_i = 0 .
\end{equation}
Using the Einstein constraint equations, this condition relates the initial
metric perturbation to the total density and velocity perturbations as
\begin{equation}
\label{eq:IC_constraints}
\Psi_i
=
-\frac{\delta_i}{2\!\left[1+k^2/(3\mathcal{H}_i^2)\right]},
\qquad
\Psi_i
=
-\frac{3}{2}\,\mathcal{H}_i\,v_i\,(1+w_{{\rm eff},i}) .
\end{equation}
Although the initial conditions are imposed in the limit $k/\mathcal{H}_i\ll 1$,
we retain the term proportional to $(k/\mathcal{H}_i)^2$ in
Eq.~\eqref{eq:IC_constraints}. This ensures that $\Psi'_i$ vanishes exactly at the
initial time and prevents the appearance of small numerical transients at the
start of the integration.

We assume adiabatic initial conditions for the fluid components. Adiabaticity
implies that the density perturbations satisfy \cite{Ballesteros:2010ks}
\begin{equation}
\label{eq:IC_adiabatic_densities}
\frac{\delta_{{\rm r}i}}{1+w_{\rm r}}
=
\frac{\delta_{{\rm m}i}}{1+w_{\rm m}}
=
\frac{\delta_i}{1+w_{{\rm eff},i}},
\end{equation}
and that all components share a common velocity potential,
\begin{equation}
\label{eq:IC_adiabatic_velocities}
v_{{\rm r}i}=v_{{\rm m}i}=v_i ,
\end{equation}
with $w_{\rm r}=1/3$ and $w_{\rm m}=0$. Combining
Eqs.~\eqref{eq:IC_constraints}–\eqref{eq:IC_adiabatic_velocities}, the initial
density and velocity perturbations are completely fixed in terms of $\Psi_i$ and
$w_{{\rm eff},i}$. At the initial time considered here, radiation domination
implies $w_{{\rm eff},i}\simeq1/3$.

In contrast to the fluid components, the scalar field is treated as a fundamental
degree of freedom rather than as an effective fluid. For numerical convenience,
we choose
\begin{equation}
\delta\phi_i = 0,
\qquad
\delta\phi'_i = 0 .
\end{equation}
These conditions do not enforce exact adiabaticity in the scalar-field sector at
the initial time. However, this choice has no impact on the physical predictions
of the model. At $a=a_i$ the scalar field is dynamically subdominant,
$\Omega_\phi(a_i)\ll 1$, so its perturbations do not contribute appreciably to the
Einstein constraints. Moreover, on super-Hubble scales during radiation
domination, the coupled Einstein–Klein–Gordon system admits an attractive
adiabatic growing mode. As a result, the evolution rapidly converges toward the
adiabatic solution before horizon entry for the modes of interest, as we show in Appendix~\ref{app:fluid_pathology_metric_regular}.

For the numerical calculations we set $\Psi_i=1$. Physical amplitudes are restored
\emph{a posteriori} by rescaling the solutions using the primordial perturbation
amplitude inferred from cosmic microwave background observations. Concretely, the
dimensionless numerical solutions are multiplied by the factor
\begin{equation}
\label{eq:IC_normalisation}
\mathcal{A}(k)
=
\frac{2\pi}{3}\,\sqrt{2A_s}
\left(\frac{k}{k_{\rm pivot}}\right)^{\frac{n_s-1}{2}} k^{-3/2},
\end{equation}
which relates the initial Newtonian potential to the primordial curvature power
spectrum generated during inflation.
Throughout this work we adopt
$\ln(10^{10}A_s)=3.0448$, $n_s=0.96605$, and
$k_{\rm pivot}=0.05\,\mathrm{Mpc}^{-1}$, motivated by the Planck 2018 results
\cite{Planck:2018vyg}.
We note that, unlike the background parameter $H_0$, which we choose to lie between
the Planck and Type~Ia supernova preferred values, the primordial parameters
$(A_s,n_s)$ are constrained by CMB anisotropies and are not informed by supernova
data. It is therefore natural to adopt the Planck-preferred values when normalising
the perturbation spectrum.

\section{Numerical results} \label{numerical results}

In this section we present the numerical solutions of the linear perturbation
equations derived in the previous section. Our aim is to characterise the
behaviour of scalar-field and matter perturbations in generalised axion-like
quintessence models, with particular emphasis on the oscillatory regime and its
impact on structure formation. We first analyse the evolution of the perturbations
themselves and then examine the resulting signatures in cosmological observables.

\subsection{Scalar-field and matter perturbations}
\label{subsec:sf_matter_perturbations}

We now analyse the numerical evolution of scalar-field and matter perturbations and
their dependence on the underlying dark-energy dynamics. This joint discussion
allows us to assess both the regularity of the perturbations in the oscillatory
regime and the impact of the scalar field on the growth of cosmic structure.

\begin{figure*}[t]
  \centering
  \begin{subfigure}[t]{0.48\textwidth}
    \centering
    \includegraphics[width=\textwidth]{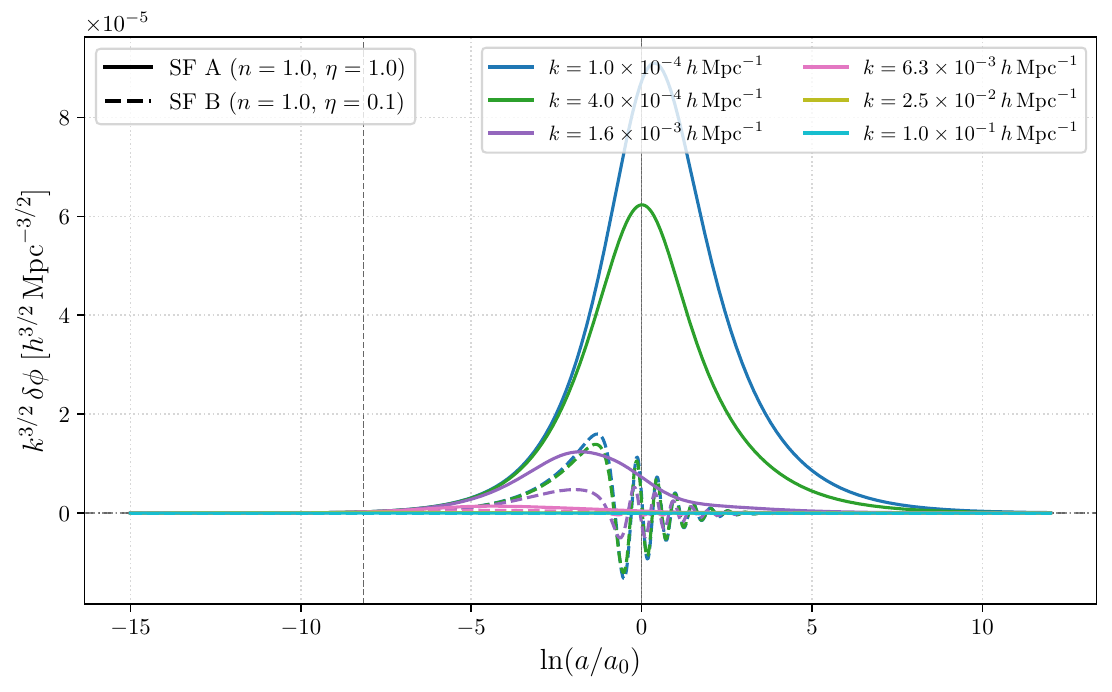}
    \caption{\justifying{Evolution of the scalar-field perturbations
    $k^{3/2}\delta\phi$ as a function of $\ln(a/a_0)$ for several Fourier modes.
    The perturbations remain finite and well behaved throughout the evolution.
    The non-oscillatory model exhibits larger amplitudes, while in the
    oscillatory case the perturbations are suppressed as the scalar-field
    equation of state rapidly approaches and oscillates around $w_\phi=-1$.}}
    \label{fig:sf_pert}
  \end{subfigure}
  \hfill
  \begin{subfigure}[t]{0.47\textwidth}
    \centering
    \includegraphics[width=\textwidth]{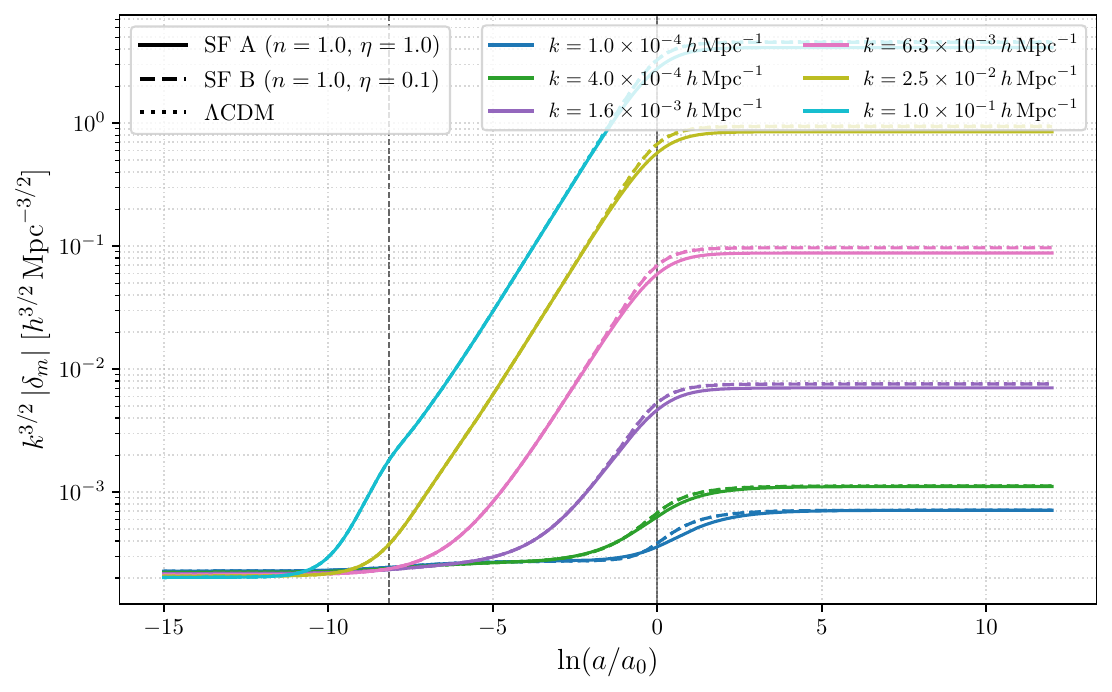}
    \caption{\justifying{Evolution of the matter density perturbations
    $k^{3/2}|\delta_m|$ for the same set of Fourier modes.
    In the non-oscillatory case a late-time suppression of matter clustering is
    observed due to the prolonged tracking regime. In contrast, in the
    oscillatory case the evolution of matter perturbations is practically
    indistinguishable from that of $\Lambda$CDM.}}
    \label{fig:matter_pert}
  \end{subfigure}
  \caption{\justifying{Scalar-field and matter perturbations for generalised
  axion-like quintessence models. The left panel shows the evolution of the
  scalar-field perturbations, highlighting their regular behaviour across the
  oscillatory regime. The right panel shows the corresponding matter density
  perturbations, illustrating the suppression present in the non-oscillatory
  case and its absence in the oscillatory case, where the dynamics closely
  resemble $\Lambda$CDM.}}
  \label{fig:sf_matter_pert}
\end{figure*}

Figure~\ref{fig:sf_matter_pert} shows the evolution of the scalar-field
perturbations (left panel) and the matter density perturbations (right panel) for
several Fourier modes, comparing a non-oscillatory model with an oscillatory one.
In all cases, the perturbations remain finite and well behaved throughout the
entire evolution. In particular, no divergences appear when the background scalar
field crosses points where $\dot{\phi}=0$. This confirms that the field-based
formulation adopted in this work provides a fully regular description of linear
perturbations, even in the presence of rapid oscillations.

The appearance of oscillations in $\delta\phi$ for the background--oscillatory
solutions can be understood from the structure of the scalar-field perturbation
equation \eqref{eq:delta_phi_KG}, which takes the form of a damped, driven oscillator for each Fourier
mode. The effective frequency is set by the combination
$k^2+a^2V_{,\phi\phi}(\bar\phi)$, while the perturbations are sourced by the metric
and background evolution through terms proportional to $\bar\phi'$ and
$V_{,\phi}(\bar\phi)$. Once the background field oscillates around the minimum of
the potential, both $\bar\phi'$ and
$V_{,\phi}(\bar\phi)\simeq m_{\rm eff}^2(\bar\phi-\phi_{\rm min})$ inherit the same
oscillatory time dependence, so that the source term becomes oscillatory and
naturally excites an oscillatory response in $\delta\phi$, even though the
perturbations remain perfectly regular throughout the evolution. In addition,
near the minimum of the potential the curvature $V_{,\phi\phi}(\bar\phi)$ is
approximately constant and equal to $m_{\rm eff}^2$, so that the homogeneous part
of the perturbation equation admits oscillatory solutions whenever this scale
dominates over Hubble damping. This situation is realised in the oscillatory
benchmark, whereas in the non-oscillatory case the scalar field remains
effectively overdamped at late-time, suppressing any coherent oscillatory pattern
in $\delta\phi$.

The amplitude of the scalar-field perturbations is larger in the non-oscillatory
case. In this regime the scalar field remains in a tracking phase until
relatively late-time, with an equation of state that deviates appreciably from
$w_\phi=-1$. During tracking, the scalar field develops comparatively larger
fluctuations, leading to a higher late-time amplitude of $\delta\phi$. In
contrast, in the oscillatory case the scalar field has a larger effective mass and
exits the tracking regime earlier. Once the scalar-field energy density becomes
relevant, its equation of state rapidly approaches $w_\phi\simeq -1$ and
subsequently oscillates around this value. As a result, the scalar-field
perturbations are suppressed and their amplitude remains small, approaching the
$\Lambda$CDM limit, where dark energy does not cluster.

A complementary behaviour is observed in the matter sector. In the oscillatory
case, the scalar field behaves effectively as a cosmological constant by late-time, and the evolution of matter perturbations is practically
indistinguishable from that of $\Lambda$CDM. Any suppression of matter clustering
is therefore negligible. In contrast, in the non-oscillatory case the prolonged
tracking phase leads to a late-time suppression of matter perturbations. Since the
scalar field contributes non-negligibly to the total energy density while
maintaining an equation of state significantly different from $-1$, the expansion
history and the gravitational potentials driving structure formation are modified,
resulting in a reduced growth rate and less efficient clustering. As we will show
in the next subsection, this suppression translates directly into observable
signatures and can be used to constrain the parameter space of the model \cite{Chiang:2025qxg}.

\subsection{Cosmological observables}
\label{subsec:cosmological_observables}

We conclude our numerical analysis by examining the impact of the scalar-field
dynamics on cosmological observables related to structure formation. In
particular, we focus on the matter power spectrum and on the quantity
$f\sigma_8$, both of which provide direct observational probes of the growth of
matter perturbations. As we show below, these observables encode the same
qualitative behaviour already identified at the level of linear perturbations.

In the Newtonian gauge adopted throughout this work, the (dimensionless) matter
power spectrum can be expressed as \cite{Albarran:2016mdu}
\begin{equation}
\label{eq:matter_power_spectrum}
P(k,z)
=
\left|\delta_m(k,z)-3\mathcal{H}(z)\,v_m(k,z)\right|^2 ,
\end{equation}
where $\delta_m$ and $v_m$ denote the matter density contrast and velocity
potential, respectively, and $\mathcal{H}$ is the conformal Hubble rate. This
quantity characterises the distribution of matter fluctuations as a function of
scale.

Another widely used observable in large-scale structure analyses is
$f\sigma_8$, which combines the growth rate of matter perturbations \cite{Balcerzak:2012ae},
\begin{equation}
\label{eq:growth_rate}
f(z) \equiv \frac{\mathrm d\ln\delta_m}{\mathrm d\ln a},
\end{equation}
with the rms amplitude of matter fluctuations on scales of
$8\,h^{-1}\,\mathrm{Mpc}$, denoted by $\sigma_8$. Its redshift evolution is given
by \cite{Wang:2010gq}
\begin{equation}
\label{eq:sigma8}
\sigma_8(z)
=
\sigma_8(0)\,
\frac{\delta_m(z,k_{\sigma_8})}{\delta_m(0,k_{\sigma_8})},
\end{equation}
where $k_{\sigma_8}=0.125\,h\,\mathrm{Mpc}^{-1}$ corresponds to a comoving scale of
$8\,h^{-1}\,\mathrm{Mpc}$ and we take $\sigma_8(0)=0.8120$ \cite{Planck:2018vyg}. The product $f\sigma_8$ is particularly useful, as it is
largely insensitive to galaxy bias and can be directly compared with
observational data.

\begin{figure*}[t]
  \centering
  \begin{subfigure}[t]{0.48\textwidth}
    \centering
    \includegraphics[width=\textwidth]{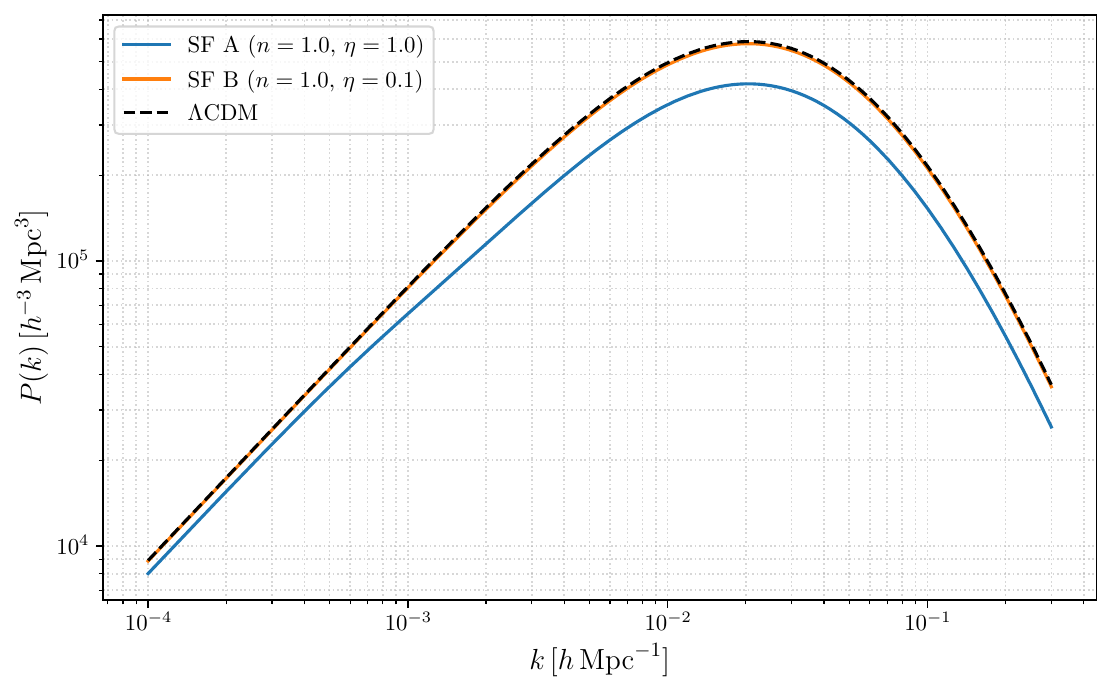}
    \caption{\justifying{Matter power spectrum at $z=0$ for the non-oscillatory
    and oscillatory benchmark models, compared with $\Lambda$CDM. A clear
    suppression is observed in the non-oscillatory case due to the prolonged
    tracking regime, while in the oscillatory case the spectrum closely follows
    the $\Lambda$CDM prediction.}}
    \label{fig:mps}
  \end{subfigure}
  \hfill
  \begin{subfigure}[t]{0.48\textwidth}
    \centering
    \includegraphics[width=\textwidth]{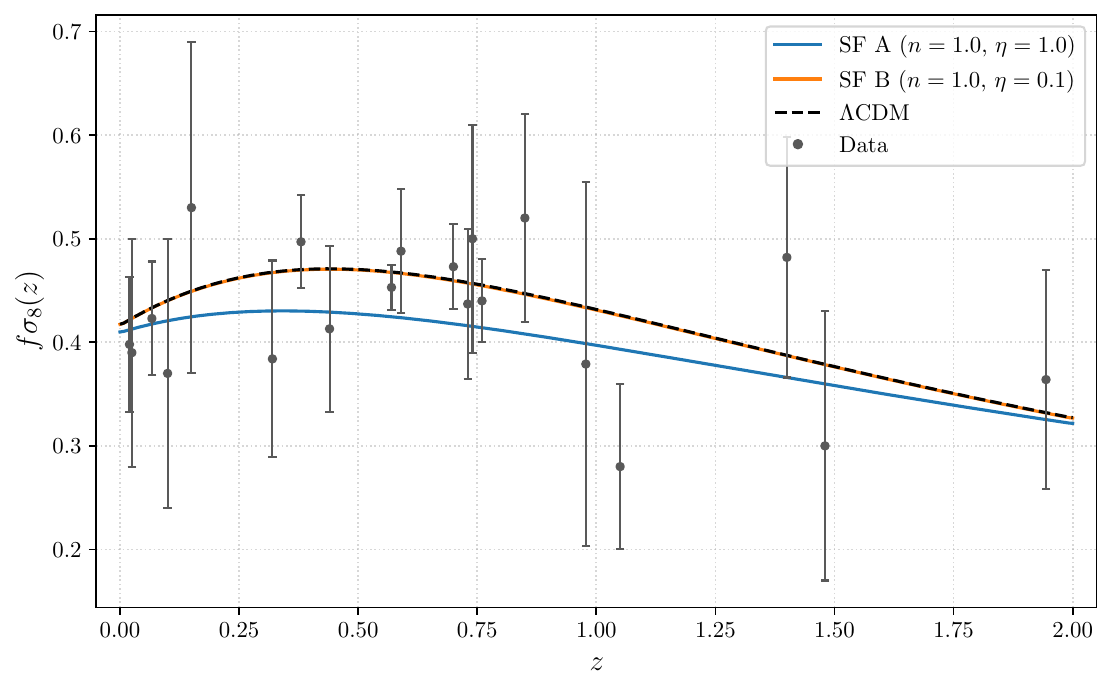}
    \caption{\justifying{Evolution of $f\sigma_8$ as a function of redshift for
    the same benchmark models, compared with $\Lambda$CDM and observational
    data \cite{Avila:2022xad}. The non-oscillatory model exhibits a noticeable suppression at low
    redshift, while the oscillatory case remains close to the $\Lambda$CDM
    prediction.}}
    \label{fig:fs8}
  \end{subfigure}
  \caption{\justifying{Cosmological observables in generalised axion-like
  quintessence models. The left panel shows the matter power spectrum at $z=0$,
  while the right panel displays the evolution of $f\sigma_8$ at low redshift.
  Both observables reflect the same qualitative behaviour found at the level of
  linear perturbations: significant suppression in the non-oscillatory case and
  near-$\Lambda$CDM behaviour in the oscillatory case.}}
  \label{fig:observables}
\end{figure*}

Figure~\ref{fig:observables} summarises our results. The left panel shows the
matter power spectrum at $z=0$ for the non-oscillatory and oscillatory benchmark
models, compared with the $\Lambda$CDM prediction. The right panel displays the
evolution of $f\sigma_8$ at low redshift, together with observational data.

In the non-oscillatory case, the matter power spectrum exhibits a clear
suppression with respect to $\Lambda$CDM over a wide range of scales. This
suppression is a direct consequence of the prolonged tracking regime discussed
in the previous subsection. Since the scalar field departs significantly from
$w_\phi=-1$ while contributing non-negligibly to the total energy density, the
growth of matter perturbations is reduced, leading to less efficient clustering
and a lower amplitude of the power spectrum at late-time. The same effect is
reflected in the evolution of $f\sigma_8$, which is systematically suppressed at
low redshift.

In contrast, in the oscillatory case the deviations from $\Lambda$CDM are
significantly reduced. Owing to the larger effective mass of the scalar field,
the tracking regime ends earlier and the equation of state rapidly approaches
$w_\phi\simeq -1$ once the scalar-field energy density becomes relevant. As a
result, the late-time expansion history and the growth of structures closely
resemble those of $\Lambda$CDM. This behaviour is clearly visible in both
observables: the matter power spectrum closely follows the $\Lambda$CDM
prediction, and the evolution of $f\sigma_8$ remains well within current
observational bounds.

Overall, these results demonstrate that structure-formation observables provide
a powerful discriminator between different scalar-field dynamics. While
non-oscillatory models with extended tracking phases are subject to strong
constraints due to their suppression of matter clustering, oscillatory models
naturally evade these bounds by approaching a cosmological-constant–like
behaviour at late-time, rendering them largely indistinguishable from
$\Lambda$CDM.

\section{Conclusions}\label{conclusions}

We have investigated late-time cosmic acceleration driven by a generalised axion-like
quintessence field in the regime in which the scalar field executes coherent oscillations
around the finite, positive minimum of its potential. This oscillatory phase is a
distinctive feature of the class of models considered here, and it has a direct impact on
the formulation of linear cosmological perturbations. In particular, we have shown that
the commonly used effective multi-fluid description of dark energy becomes ill defined
once oscillations occur: although the background quantities remain regular, fluid
variables such as the adiabatic sound speed and combinations involving $1+w_\phi$ develop
turning-point divergences when the field reaches $w_\phi=-1$ at each oscillation.

To address this issue, we developed and implemented a field-based perturbation framework
that evolves the fundamental variables $(\Psi,\delta\phi)$ together with standard matter
and radiation perturbations in the Newtonian gauge. This formulation remains fully
regular across the entire evolution, including at the turning points where $\dot\phi=0$,
because the Einstein equations are sourced only by the regular combinations
$\delta\rho$, $\delta p$, and $(\bar\rho+\bar p)v$. As a result, the metric perturbation
$\Psi$ and all physical perturbations remain smooth even when the mapping to fluid
variables breaks down. This provides a consistent and robust route to study structure
formation in oscillating quintessence models without relying on ill-defined effective
fluid quantities.

Using numerical solutions, we characterised the behaviour of scalar-field and matter
perturbations and connected them to late-time observables. We found that the amplitude
and phenomenology depend strongly on whether the model is non-oscillatory or oscillatory
by today, which is controlled by the comparison between the effective mass near the
minimum and the Hubble scale. In non-oscillatory solutions, the scalar field can remain
in a prolonged tracking phase with $w_\phi$ appreciably different from $-1$, leading to
a noticeable late-time suppression of matter clustering. This suppression propagates to
structure-formation observables, yielding a reduced matter power spectrum amplitude and
a systematically lower $f\sigma_8$ at low redshift relative to $\Lambda$CDM. In
oscillatory solutions, instead, the field approaches a cosmological-constant--like
behaviour once it becomes dynamically relevant and subsequently oscillates around
$w_\phi\simeq -1$, which suppresses scalar-field perturbations and makes the growth of
matter perturbations and the associated observables closely track the $\Lambda$CDM
predictions.

Overall, our results show that oscillatory generalised axion-like quintessence can
naturally evade late-time growth constraints that are more restrictive for non-oscillatory
tracking-like models, while still providing a dynamical origin for dark energy. Beyond
this qualitative conclusion, the main outcome of this work is methodological: the
field-based perturbation treatment derived here supplies a unified, well-defined
framework to explore the full parameter space of these models, including regions in
which fluid-based approaches fail.


\appendix

\section{Fluid-pathology diagnostics and regularity of the metric perturbation}
\label{app:fluid_pathology_metric_regular}

The oscillatory regime discussed in the main text highlights a well-known limitation
of the effective multi-fluid description of a canonical scalar field.
In particular, during coherent oscillations around the minimum of the potential,
the background equation of state $w_\phi$ periodically reaches $-1$ at the turning
points of the motion, i.e.\ when $\bar\phi'=0$.
While the fundamental perturbation variables $(\delta\phi,\delta\phi')$ remain regular,
several \emph{fluid} quantities commonly used in the multi-fluid framework
become ill-defined at those instants. This appendix provides a numerical diagnostic
of this pathology and shows explicitly that it does \emph{not} propagate to the metric
perturbations, which remain perfectly regular.

\vspace{0.1cm}
\noindent
\textbf{Scalar-field perturbations in field variables.}
In the Newtonian gauge and in conformal time, the scalar-field energy-density,
pressure, and momentum-density perturbations can be written directly in terms of
$\delta\phi$ as
\begin{eqnarray}
\label{eq:app_deltarho_phi}
&&\delta\rho_\phi
=
\frac{1}{a^2}\Big(
\bar\phi'\,\delta\phi'
-\bar\phi'^2\,\Psi
\Big)
+V_{,\phi}(\bar\phi)\,\delta\phi,\\[0.15cm]
\label{eq:app_deltap_phi}
&&\delta p_\phi
=
\frac{1}{a^2}\Big(
\bar\phi'\,\delta\phi'
-\bar\phi'^2\,\Psi
\Big)
-V_{,\phi}(\bar\phi)\,\delta\phi,\\[0.15cm]
\label{eq:app_momentum_phi}
&&(\bar\rho_\phi+\bar p_\phi)\,v_\phi
=
-\frac{\bar\phi'}{a^2}\,\delta\phi,
\end{eqnarray}
where an overbar denotes a background quantity and
$V_{,\phi}\equiv \mathrm{d}V/\mathrm{d}\phi$.
The background density and pressure are
\begin{equation}
\bar\rho_\phi=\frac{\bar\phi'^2}{2a^2}+V(\bar\phi),
\qquad
\bar p_\phi=\frac{\bar\phi'^2}{2a^2}-V(\bar\phi),
\qquad
w_\phi\equiv\frac{\bar p_\phi}{\bar\rho_\phi}.
\end{equation}
Therefore,
\begin{equation}
\label{eq:app_one_plus_w}
1+w_\phi=\frac{\bar\rho_\phi+\bar p_\phi}{\bar\rho_\phi}
=\frac{\bar\phi'^2}{a^2\,\bar\rho_\phi},
\end{equation}
so that $1+w_\phi=0$ precisely when $\bar\phi'=0$, i.e.\ when $w_\phi$ reaches $-1$.

\vspace{0.1cm}
\noindent
\textbf{A diagnostic of the multi-fluid pathology.}
A commonly used ``adiabatic-like'' combination in the fluid treatment is
$\delta_\phi/(1+w_\phi)$, where $\delta_\phi\equiv\delta\rho_\phi/\bar\rho_\phi$.
Using \eqref{eq:app_deltarho_phi}--\eqref{eq:app_one_plus_w}, we can write
\begin{equation}
\label{eq:app_delta_over_1pw}
\frac{\delta_\phi}{1+w_\phi}
=
\frac{\delta\rho_\phi/\bar\rho_\phi}{\bar\phi'^2/(a^2\bar\rho_\phi)}
=
\frac{\bar\phi'\,\delta\phi'
-\bar\phi'^2\Psi
+a^2V_{,\phi}\delta\phi}{\bar\phi'^2}.
\end{equation}
Even though $\delta\rho_\phi$, $\delta p_\phi$, and $(\bar\rho_\phi+\bar p_\phi)v_\phi$
remain finite, the ratio \eqref{eq:app_delta_over_1pw} diverges whenever $\bar\phi'=0$
(equivalently when $w_\phi$ reaches $-1$). The same turning-point pathology underlies
the divergence of other fluid variables, such as the scalar-field velocity potential
\begin{equation}
\label{eq:app_vphi}
v_\phi=-\frac{\delta\phi}{\bar\phi'},
\end{equation}
and the adiabatic sound speed
$c_{a,\phi}^2\equiv \bar p_\phi'/\bar\rho_\phi'$.
Importantly, these divergences are \emph{kinematical} and do not signal any physical
instability: they simply reflect the breakdown of the mapping from field variables
to a fluid description at the turning points of the background evolution.

\vspace{0.1cm}
\noindent
\textbf{Metric equations remain regular.}
In the Einstein equations, the metric perturbation is sourced by the \emph{total}
density perturbation, momentum density, and pressure perturbation. In the notation
used in the main text (total quantities without subindices), the sources that enter
the linearised Einstein equations are
\begin{eqnarray}
\label{eq:app_total_sources}
&&\delta\rho = \delta\rho_{\rm m} + \delta\rho_{\rm r} + \delta\rho_\phi,\\
&&\delta p   = \delta p_{\rm r} + \delta p_\phi,\\
&&(\bar{\rho}+\bar{p})\,v
= \sum_A(\bar{\rho}_A+\bar{p}_A)\,v_A.
\end{eqnarray}
For matter and radiation, all contributions are manifestly regular.
For the scalar field, $\delta\rho_\phi$, $\delta p_\phi$, and
$(\bar\rho_\phi+\bar p_\phi)v_\phi$ are given by
Eqs.~\eqref{eq:app_deltarho_phi}--\eqref{eq:app_momentum_phi} and remain finite
throughout the oscillatory regime.
Hence, even though fluid variables such as
$\delta_\phi/(1+w_\phi)$, $v_\phi$, and $c_{a,\phi}^2$
diverge at turning points, the combinations that actually source the metric do not,
and the gravitational potential $\Psi$ remains well behaved.

\begin{figure*}[t]
  \centering
  \begin{subfigure}[t]{0.48\textwidth}
    \centering
    \includegraphics[width=\textwidth]{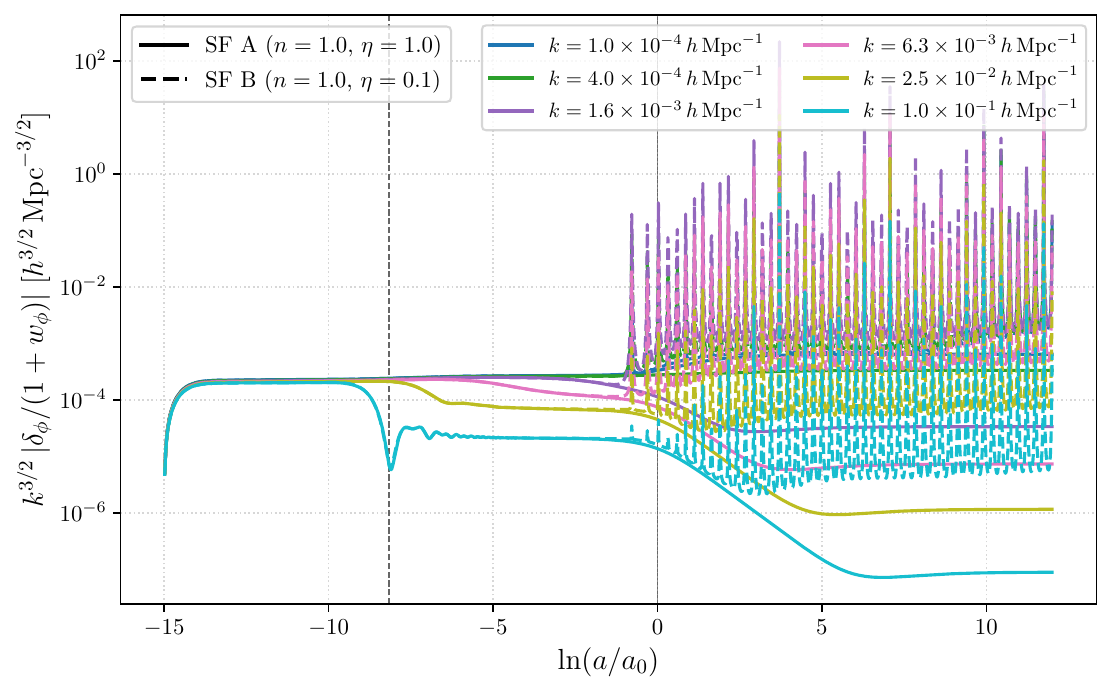}
    \caption{\justifying{Evolution of the ``adiabatic-like'' fluid diagnostic
    $k^{3/2}\,|\delta_\phi/(1+w_\phi)|$ for the benchmark models. The divergent spikes
    occur only in the oscillatory benchmark and appear whenever the background equation
    of state reaches $w_\phi=-1$ (equivalently $\bar\phi'=0$). The locations of the peaks
    are independent of the mode $k$, as expected since the turning points are fixed by
    the background dynamics. In the non-oscillatory benchmark no such spikes arise.}}
    \label{fig:app_fluid_pathology}
  \end{subfigure}
  \hfill
  \begin{subfigure}[t]{0.48\textwidth}
    \centering
    \includegraphics[width=\textwidth]{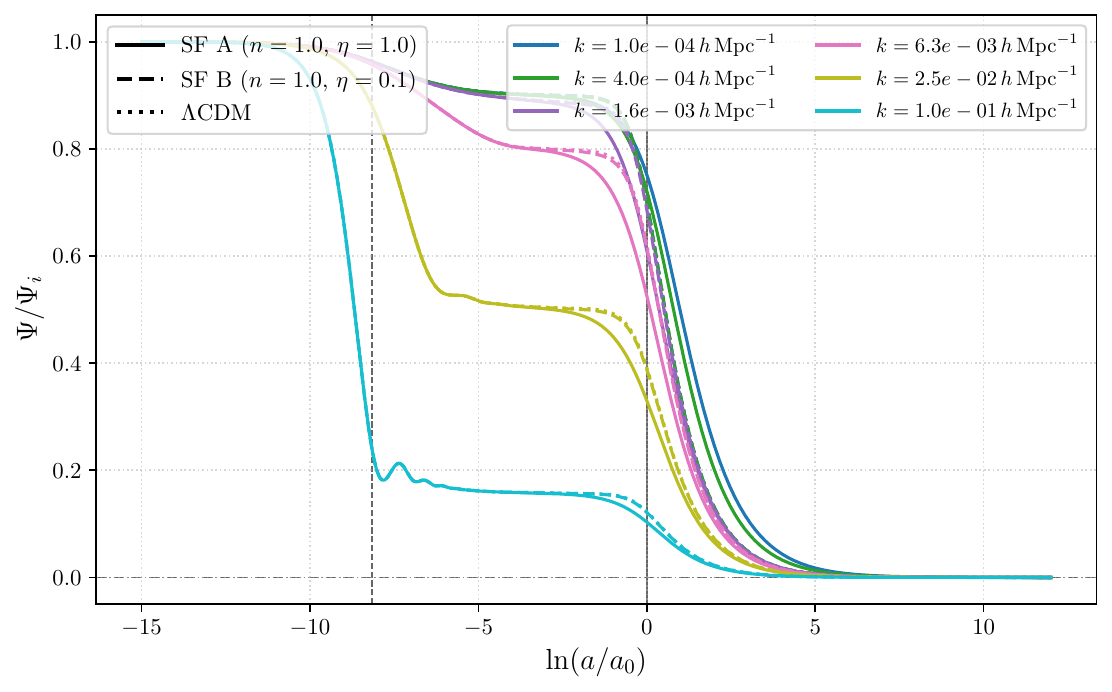}
    \caption{\justifying{Evolution of the metric potential $\Psi/\Psi_i$ for the same
    set of modes and models (non-oscillatory and oscillatory scalar-field benchmarks,
    and $\Lambda$CDM). No pathology is observed: $\Psi$ remains finite and smooth
    throughout the oscillatory regime because the Einstein equations are sourced by the
    regular combinations $\delta\rho$, $(\bar\rho+\bar p)\,v$, and $\delta p$
    (cf.~\eqref{eq:app_total_sources}), rather than by ill-defined fluid variables such
    as $\delta_\phi/(1+w_\phi)$, $v_\phi$, or $c_{a,\phi}^2$.}}
    \label{fig:app_metric_regular}
  \end{subfigure}
  \caption{\justifying{Diagnostic of the breakdown of the multi-fluid description in
  the oscillatory regime and regularity of the metric perturbation. The left panel shows
  that the fluid diagnostic $\delta_\phi/(1+w_\phi)$ develops turning-point divergences
  only for the oscillatory benchmark. The right panel shows that this behaviour does not
  propagate to the metric: the gravitational potential remains well defined and well
  behaved, and closely follows the corresponding smooth evolution in the non-oscillatory
  benchmark and in $\Lambda$CDM.}}
  \label{fig:app_fluid_pathology_metric}
\end{figure*}

The behaviour discussed above is illustrated in Fig.~\ref{fig:app_fluid_pathology_metric}.
The left panel shows the fluid-pathology diagnostic $\delta_\phi/(1+w_\phi)$ for several
Fourier modes, including both benchmark scalar-field models (non-oscillatory and
oscillatory), and the $\Lambda$CDM case for comparison. Divergent spikes appear
\emph{only} in the oscillatory benchmark and occur at the same values of
$\ln(a/a_0)$ for all modes, confirming that they are driven by background turning
points at which $w_\phi$ reaches $-1$, rather than by any scale-dependent instability.
All the divergent peaks coincide with the minima of the oscillations in the
scalar-field equation of state shown in Fig.~\ref{fig:background_wphi}.

Although the majority of these turning points occur in the future for the benchmark
oscillatory model shown here, some of them already take place close to the present
epoch, making their treatment relevant for late-time cosmology. Moreover, models
with larger effective masses—e.g.\ obtained by considering smaller values of
$\eta$—exhibit more frequent oscillations, with multiple turning points occurring
well before today. In all such cases, the standard fluid description becomes ill
defined, independently of whether the turning points lie in the past or future.
This motivates the use of the field-based perturbation framework developed in this
work, which provides a consistent description across the full parameter space,
including models with heavier fields and earlier oscillatory behaviour.

Finally, although the scalar-field perturbations are
initialised with $\delta\phi_i=\delta\phi_i'=0$, which does not strictly satisfy
adiabatic initial conditions, the evolution rapidly converges to the adiabatic attractor
on super-Hubble scales, as
is clearly visible in the early-time behaviour of the fluid diagnostic.

The right panel shows the evolution of the metric potential $\Psi/\Psi_i$ for the same
set of modes and models. Despite the fluid-pathology spikes present in the oscillatory
scalar-field benchmark, $\Psi$ remains finite and smooth at all times. This confirms that
the apparent divergences arising in the multi-fluid description do not propagate to the
metric sector, since the Einstein equations are sourced only by the regular combinations
$\delta\rho$, $(\bar\rho+\bar p)\,v$, and $\delta p$.

\section*{Acknowledgements}
 The authors are grateful to Hsu-Wen Chiang for his helpful comments on this work.
 M. B.-L. is supported by the Basque Foundation of Science Ikerbasque. C. G. B. acknowledges financial support from the FPI fellowship PRE2021-100340 of the Spanish Ministry of Science, Innovation and Universities.  Our work is supported by the Spanish Grant PID2023-149016NB-I00 (MINECO/AEI/FEDER, UE).  This work is also supported by the Basque government Grant No. IT1628-22 (Spain). 

\bibliography{bibliografia.bib}

\end{document}